# An Availability Management Framework for Microservice-Based Safety-Critical CIoT Systems – Architecture and End-to-End Evaluation

Hassaan Siddiqui, and Ferhat Khendek, *Member, IEEE*

***Abstract*—Safety-critical Cellular Internet of Things (CIoT) systems increasingly rely on microservice architectures orchestrated across heterogeneous edge and cloud platforms. Although this paradigm improves modularity and scalability, the fault-tolerance mechanisms of contemporary orchestration platforms remain predominantly reactive, i.e., failed components are restarted or rescheduled only after an outage has already occurred. Such recovery can produce multi-second End to End Service Outage Time (E2E SOT), which can grow further under container restart backoff. This paper proposes an Availability Management Framework (AMF), a generic monitor-predict-recover loop that complements reactive recovery with machine learning-based anomaly detection and make-before-break switchover to a standby replica. Candidate ML models are evaluated offline under three protocols, including a change of operating regime, in order to select a model for live deployment. AMF is instantiated and evaluated on a three-tier tele-operated driving (ToD) testbed under a staged resource-exhaustion scenario. Relative to Kubernetes-only recovery, AMF completes switchover during the anomaly window, reduces E2E SOT, and avoids the tens-of-seconds outages associated with repeated in-place restarts under backoff. On this non-real-time testbed, residual E2E SOT lies inside healthy-path jitter and is therefore limited by path timing rather than by lateness of the recovery decision. The paper whows that coupling a validated ML based anomaly detector with proactive switchover can remove restart- and backoff-dominated outage from a critical path.**

## I. INTRODUCTION

ENSURING high availability (a.k.a as fine 9’s) has long been a central concern in dependable computing. In monolithic deployments, tightly coupled components share a common failure domain: a fault in one part of the application forces restart or redeployment of the whole system, which prolongs downtime and complicates recovery. The difficulty is amplified in safety-critical Cellular Internet of Things (CIoT) settings [1], where availability is not only a quality attribute but a prerequisite for safe operation. Such systems typically span heterogeneous platforms, including devices, edge nodes, on-premises servers, and cloud environments. They operate in contexts where sustained disruption may endanger people, damage infrastructure, or cause irreversible loss of operational data [2], [3]. Continuous End-to-End (E2E) service delivery is therefore essential.

Microservices Architecture (MA) has become a widely adopted alternative for building large distributed applications [4], [5], [6], [7]. By decomposing a system into independently deployable, loosely coupled services [8], MA improves fault isolation, incremental evolution, and elastic scaling, and thereby offers a more favorable foundation for availability than a monolithic design. At the same time, it introduces a denser failure surface. Services interact across nodes and networks; dependency failures, resource contention, and orchestration delays can cascade into observable service outages even when individual components appear healthy. Maintaining availability under these conditions requires not only platform-level orchestration, but also availability management mechanisms that can act before a fault becomes an E2E disruption.

Container orchestrators such as Kubernetes (K8s) provide automated placement, scaling, and recovery for microservice workloads, and have therefore been a natural substrate for CIoT deployments. In prior work we examined microservice-based Tele-operated Driving (ToD) services on K8s and showed that, although the platform restores failed containers, its recovery remains predominantly reactive and does not, by itself, meet the availability expectations of safety-critical CIoT applications [1], [9]. In particular, recovery begins after the failure has already interrupted the service path, so the duration of the outage is governed by detection latency, container restart, and related control-plane delays. For applications that must sustain a continuous command or control stream, such interruptions cannot be tolerated.

These observations motivate availability management mechanisms that complement reactive recovery with proactive Fault-Tolerance (FT). Machine learning (ML)–based anomaly detection offers a practical means to that end, by continuously observing resource-usage telemetry, a trained model can identify anomalous behavior that precedes potential failures and thereby create a window in which protective actions can

This paragraph of the first footnote will contain the date on which you submitted your paper for review, which is populated by IEEE. This work has been partially supported by the Natural Sciences and Engineering Research Council of Canada (NSERC). *(Corresponding author: Ferhat Khendek).*
Hassaan Siddiqui is with Tecsys Inc., Montreal, Canada and the Department of Electrical and Computer Engineering, Concordia University, Montreal, Canada (e-mail: hassaan.a.s@gmail.com).
Ferhat Khendek is with the Department of Electrical and Computer Engineering, Concordia University, Montreal, Canada (e-mail: ferhat.khendek@concordia.ca).

be taken. Building on this idea, this paper proposes an Availability Management Framework (AMF) for microservice-based safety-critical CIoT systems. The proposed AMF integrates monitoring, ML-based prediction, and recovery so that impending failures can be anticipated and mitigated, through switchover to a standby replica, before they materialize as service outages. Beyond describing the framework, in this article, we instantiate it on a three-tier ToD testbed, select and validate the deployed detector offline, and evaluate E2E service availability under a staged resource-exhaustion scenario against K8s' native recovery.

### A. Problem Statement

Safety-critical CIoT systems require continuous E2E service availability. Microservices coupled with container orchestration platforms provide automated recovery, yet that recovery is predominantly reactive, i.e., corrective action begins only after a component has already failed, so the service path may already be disrupted. Proactive FT based on ML-based anomaly detection is a promising complement, but two difficulties remain. First, an ML model must be selected and validated so that it is suitable for live use, i.e., validation under a single operating regime may overstate field performance and lead to an inappropriate deployment choice. Second, anomaly detection alone does not restore service, unless anomaly scores are consumed by a recovery action, such as switchover to a standby replica, before failure materializes, E2E availability may be no better than under reactive recovery. The overarching problem is therefore to design an availability management framework that integrates monitoring, ML-based anomaly detection, and proactive recovery in addition to the reactive one, and to determine empirically whether that approach reduces E2E service disruption relative to platform recovery alone under a representative failure scenario.

### B. Contributions

The main contributions of this work are as follows.

1. A generic AMF that combines resource monitoring, ML-based anomaly detection, and recovery actions in addition to the reactive one, including on-demand standby scale-up and make-before-break switchover, into a monitor-predict-recover loop that complements reactive orchestration for microservice-based safety-critical CIoT systems.
2. An offline ML model selection and validation procedure for the prediction module, including complementary evaluation protocols that inform which ML model is suitable for live deployment, a more detailed robustness study of these protocols appears in our prior paper [10].
3. An experimental validation of the proposed AMF on a heterogeneous three-tier ToD testbed and an evaluation of the E2E service outage under reactive recovery and under AMF-driven switchover for a staged anomaly-to-failure scenario.

A preliminary architecture proposal was presented in [11]. The present article extends that work with a concrete deployment and experimental instance, ML model selection for deployment, and E2E availability experiments and results. The remainder of the paper is organized as follows. Section II provides some background and reviews related work. Section III presents the proposed AMF. Section IV describes the ToD testbed and the deployment of the AMF.. Section V summarizes an offline ML model evaluation and selection process. Section VI presents the availability experiments and the results. Section VII discusses the findings of the paper and the threats to validity. We conclude in Section VIII.

## II. Background and Related Work

MA views applications as small, independently deployable services that encapsulate distinct functions. Together with containerization, this model improves scalability, modularity, and evolution of individual components without redeploying the entire application [12]. Container orchestrators are commonly used to operate such systems at scale. K8s, in particular, automates placement, scaling, load balancing, and basic fault recovery by monitoring container health and replacing failed units. The combination of MA and K8s therefore provides a flexible substrate for distributed, safety-critical CIoT deployments [13]. In our earlier survey of microservices for IoT [6], we concluded that MA is well suited to heterogeneous and dynamic edge-to-cloud environments [4], [5], [7], [14], [15], [16], [17], [18]. Nevertheless, high availability remains difficult to assure, each service runs in its own failure domain, dependencies can propagate faults across the chain, and orchestrator recovery is typically triggered only after a disruption has already occurred. Failures may arise from container crashes, resource exhaustion, network impairment, or memory-related faults, among other causes. In prior studies of a safety-critical CIoT testbed built on MA and K8s, we showed that reactive recovery alone does not meet the availability requirements of such applications when evaluated on E2E service behavior [1], [9].

FT mechanisms are often classified as reactive or proactive. Reactive approaches detect failures after occurence and then restore the affected component, for example, through restart or rescheduling. Proactive approaches aim to anticipate impending faults and initiate protective action earlier, thereby limiting or avoiding service interruption [19]. For microservice-based systems, several research directions pursue ML-assisted proactive FT. Tran et al. [19] combine recurrent failure prediction with proactive service migration for containerized workloads. Almaraz-Rivera [20] presents a cloud-native MLOps pipeline for anomaly-based monitoring of on-premise Kubernetes clusters, combining metric collection with Chaos Engineering to validate outlier alerts. Their focus is detection and alerting for ephemeral container workloads, rather than binding anomaly scores to proactive recovery or measuring E2E service outage time. Anemogiannis et al. [21] propose a K8s monitoring framework that builds a graph representation of cluster resources and uses graph features, together with supervised models, to highlight abnormal components in a dynamic environment. Like other detector-centric approaches, their evaluation targets abnormality identification and operator-facing visualization rather than closed-loop recovery or E2E service outage time. Matsuo and Ikegami [22] study anomaly detection in the presence of K8s auto-scaling and self-healing, showing that recovery delay depends on the abnormality type and that anomaly scores can help operators gauge the degree of degradation. Their work informs operator-facing analysis of platform self-healing rather than a proactive switchover loop that acts before failure to reduce E2E service outage time. Jin et al. [23] apply robust principal component analysis for anomaly detection and root-cause analysis in microservice environments. Raeiszadeh et al. [24] exploit distributed tracing with graph-based learning to identify anomalous behavior across service dependencies. Jia et al. [25] address stateful recovery at the industrial edge through checkpointing and log replay integrated with K8s. Complementary monitoring pipelines have been proposed that ingest Prometheus metrics and raise alerts from detected abnormalities [26], while benchmark datasets and ensemble classifiers have shown strong pod-failure prediction under stable operating conditions [27]. Vayghan et al. addressed state replication for stateful microservices through a state controller that supports high-availability recovery in K8s [13]. These contributions demonstrate that predictive monitoring and specialized recovery mechanisms can improve resilience at the service or platform level.

Despite this progress, much of the literature still evaluates success primarily through detector-centric metrics, classification accuracy, precision, recall, or F1-score, or through recovery of an individual microservice, rather than through sustained availability of an E2E service path. In safety-critical CIoT, however, the user-visible outcome is continuity of the complete service chain across devices, edge nodes, and backend platforms. A timely anomaly score on one pod does not, by itself, establish improved availability if recovery begins only after crash, if actuation is absent, or if switchover cannot complete within the precursor window. Conversely, measuring outage duration alone does not justify which detector or training protocol should be deployed in a live management loop. The gap that motivates the present study is therefore twofold; availability management must integrate monitoring, ML-based prediction, and recovery into a closed loop, and that loop must be evaluated with system-level measures of E2E service outage, not with offline classifier scores alone.

The framework proposed and evaluated in this article follows the closed control-loop tradition of autonomic computing [28], in which monitored system state drives analysis and corrective actions without operator intervention. Relative to prior detector-focused and service-level recovery studies, we emphasize offline validation of the deployed anomaly detector together with online measurement of E2E service disruption under reactive platform recovery and under proactive switchover on a three-tier testbed.

## III. The Availability Management Framework

The proposed AMF is designed to strengthen the resilience of MA-based safety-critical CIoT systems, taking into account their distributed and heterogeneous character and the limitations of MA when coupled with a container orchestrator such as K8s. The framework addresses both reactive and proactive fault management. Reactively, it supports recovery from unforeseen failures by leveraging orchestration mechanisms to restore failed microservices. Proactively, it employs anomaly detection to anticipate impending faults from observed behavioral patterns and to trigger switchover before a disruption materializes on the E2E service path. Together, these capabilities aim to sustain continuous service availability in line with the stringent availability requirements of safety-critical environments. As illustrated in Fig 1, the AMF comprises three modules: the monitoring module, the prediction module, and the recovery module.

### *A. Monitoring Module*

The monitoring module comprises two components: a healthcheck and a resource monitor. The healthcheck periodically verifies that each microservice is operating as intended. It differs from K8s default check, which primarily indicates whether a pod is running or not. The AMF healthcheck additionally assesses whether the application logic and services executing inside the pod are up and responding. This distinction matters because a pod may remain *Running* while internal code or an in-pod service misbehaves in a way that degrades the microservice's delivered function. The corresponding notification flow is; when an unexpected behavior is detected and the healthcheck fails, the recovery module is notified and initiates an appropriate failover to restore the affected microservice.

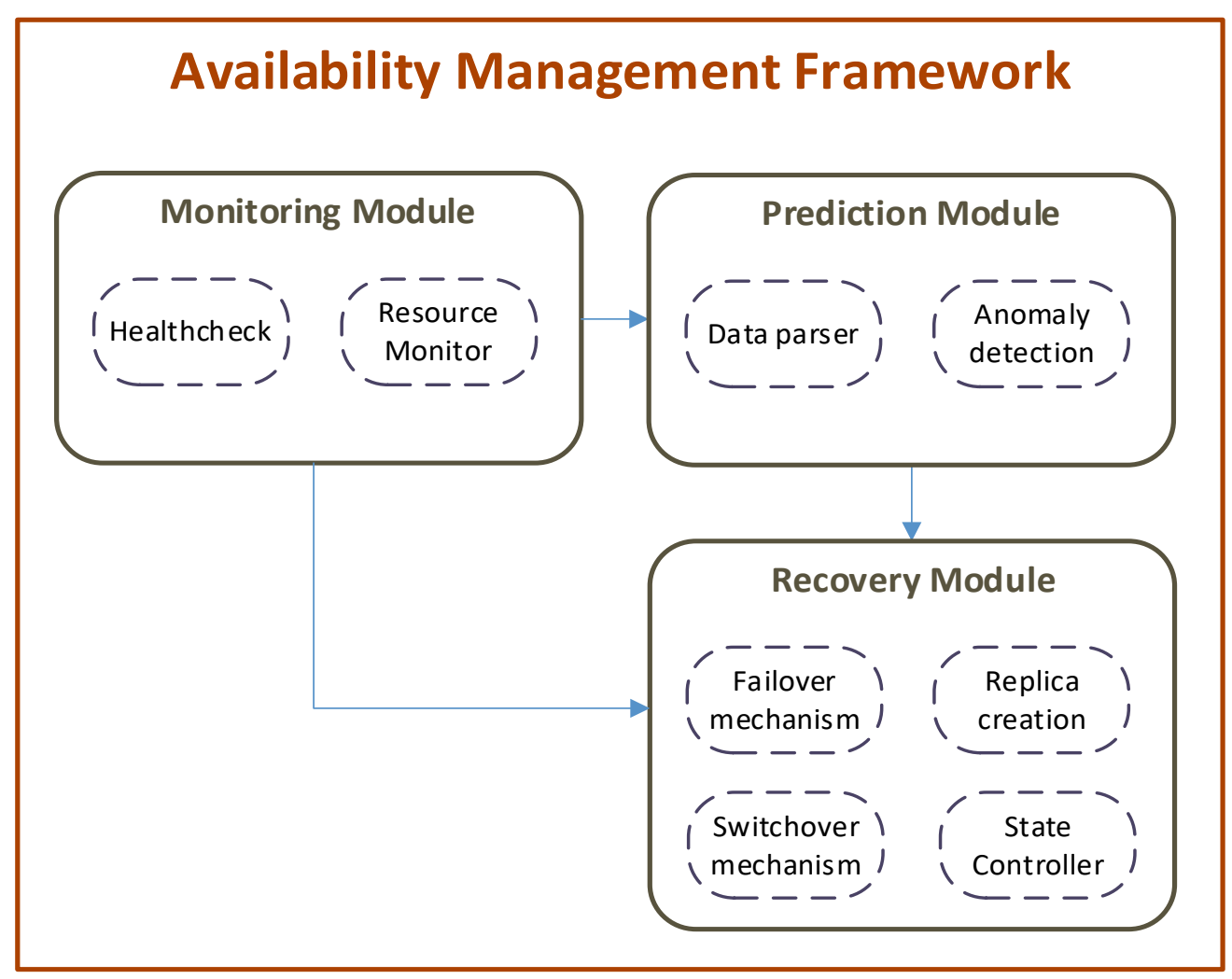

*Fig 1. The AMF architecture*

The second component of the monitoring module is the resource monitor, which serves two roles. First, it periodically collects resource-utilization metrics for each microservice, including memory and CPU usage, and stores them in an internal database. These historical records are subsequently used to train the ML model for anomaly detection. Second, after each collection interval, the resource monitor reports the current utilization values to the prediction module, which assesses whether the observed behaviour is anomalous.

### *B. Prediction Module*

The prediction module comprises two components: a data parser and an anomaly detector. The data parser retrieves records from the resource-monitor database and prepares them for consumption by the ML model. The anomaly detector follows the conventional offline-training and online-inference paradigm, as shown in Fig 2. The model is trained offline on data collected from the CIoT system and is then used to detect anomalies online. During offline training, historical telemetry is collected, parsed, and appropriately labeled. A suitable learning algorithm is trained on the labeled data, validated on held-out samples, and packaged for deployment. Because of the associated computational cost, this phase is performed offline in a controlled environment. Once deployed, the online phase operates in real time: live system data are ingested for inference without further training, and the model continuously analyses incoming streams to flag potential faults.

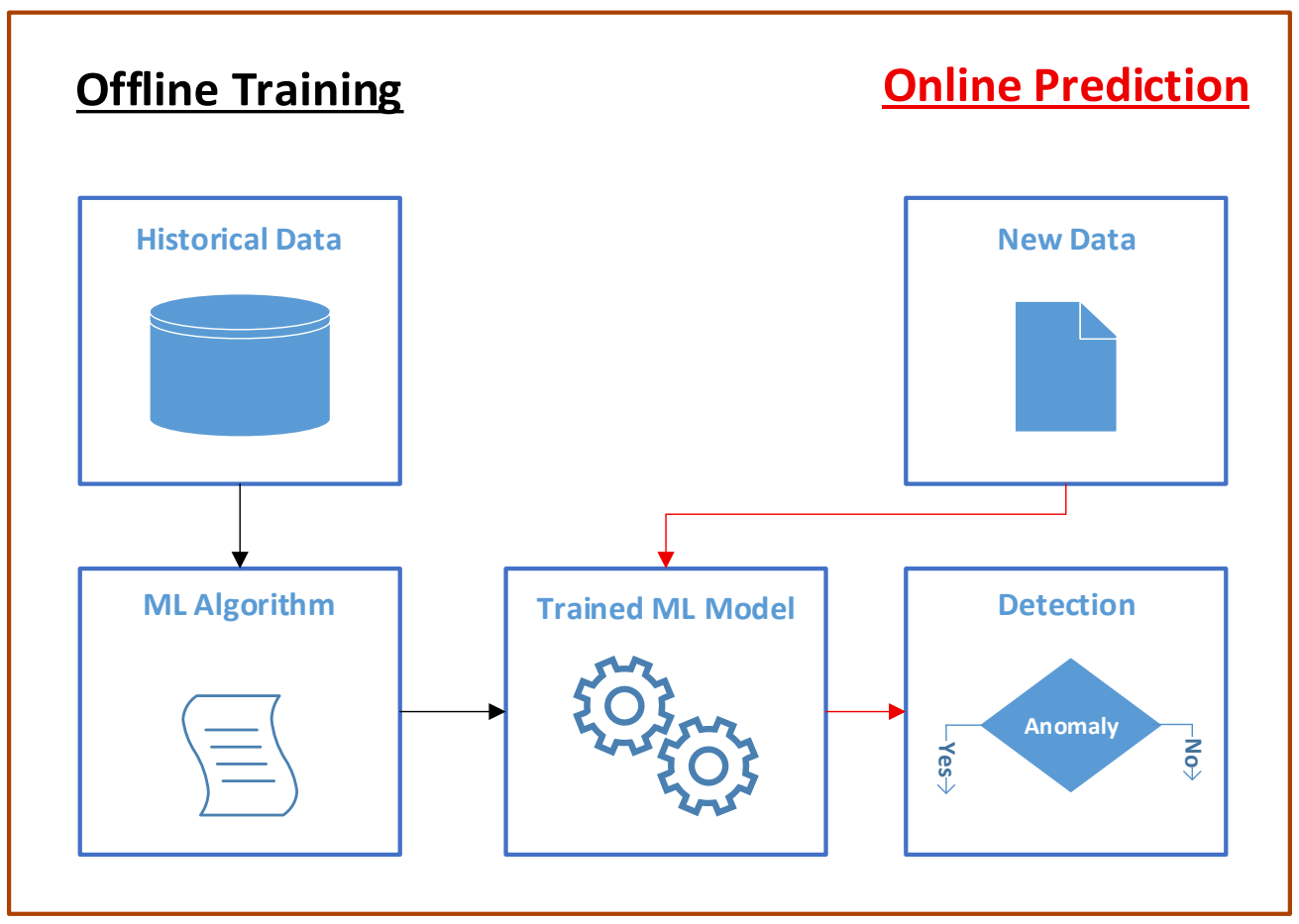

*Fig 2. ML model training vs prediction*

Selecting and training an ML model that detects anomalies accurately and precisely remains non-trivial. It requires an understanding of the data characteristics, the anomaly types of interest, and the performance requirements of the detector. In this work, the focus is on anomalies associated with elevated memory and CPU usage, which may indicate issues such as memory leaks or inefficient execution. For these anomaly types, both supervised and unsupervised learning are applicable. When the task is formulated as binary classification of normal versus anomalous instances, supervised methods such as Decision Tree or Random Forest classifiers may be used. These approaches rely on a labeled offline dataset in which normal and anomalous states are explicitly annotated. A Decision Tree partitions the feature space into subsets according to feature values, yielding a tree-structured decision model, whereas a Random Forest aggregates multiple trees to improve predictive performance. Unsupervised clustering methods are also applicable and do not require labels. Instead, they identify structure in the data by grouping similar points, and treat poorly fitting points as anomalies. Examples include K-Means, which partitions data into K clusters by feature similarity, and Fuzzy C-Means. Both families of methods have merits for detecting high memory and high CPU anomalies, the choice depends on the availability of labeled data and on the specific requirements of the detection task.

Beyond the choice of learning paradigm, the prediction module must also be validated under conditions representative of the deployment. In CIoT environments, resource-usage telemetry may change with workload and operating regime, so a detector that performs well on an offline training corpus may exhibit reduced recall or an altered precision–recall trade-off once deployed. Because AMF couples anomaly scores to recovery actions that affect E2E service availability, such degradation is not merely a classification concern, i.e., missed precursors can delay or prevent proactive switchover. Accordingly, candidate ML models should be assessed under complementary validation protocols prior to deployment, the procedure used in this study is discussed in Section V and based on our earlier offline evaluation [10].

### *C. Recovery Module*

The third module of AMF is the recovery module, which maintains service continuity by managing microservice recovery in response to detected anomalies or actual failures. It comprises four components: failover mechanisms, switchover mechanisms, replica creation, and a state controller. When a microservice fails, the system can trigger failover by redirecting its workload to a pre-existing redundant replica or by restarting the microservice, thereby limiting downtime. Alternatively, when the anomaly detector reports abnormal behavior indicative of an impending failure, the recovery module can proactively create a redundant replica of the affected microservice and switchover operation to that healthy replica before failure materializes, thereby avoiding interruption on the service path, as shown in Fig 3. However, if the abnormal behavior is not detected, failure happens and the recovery follows failure path. The choice of failover or switchover strategy, including cold standby, hot standby, or dynamic replica creation, can be configured according to the requirements of the safety-critical CIoT system, available resources, and whether the microservice is stateful or stateless. This modular and adaptive design aims to provide robust FT while remaining flexible with respect to deployment constraints and operational priorities.

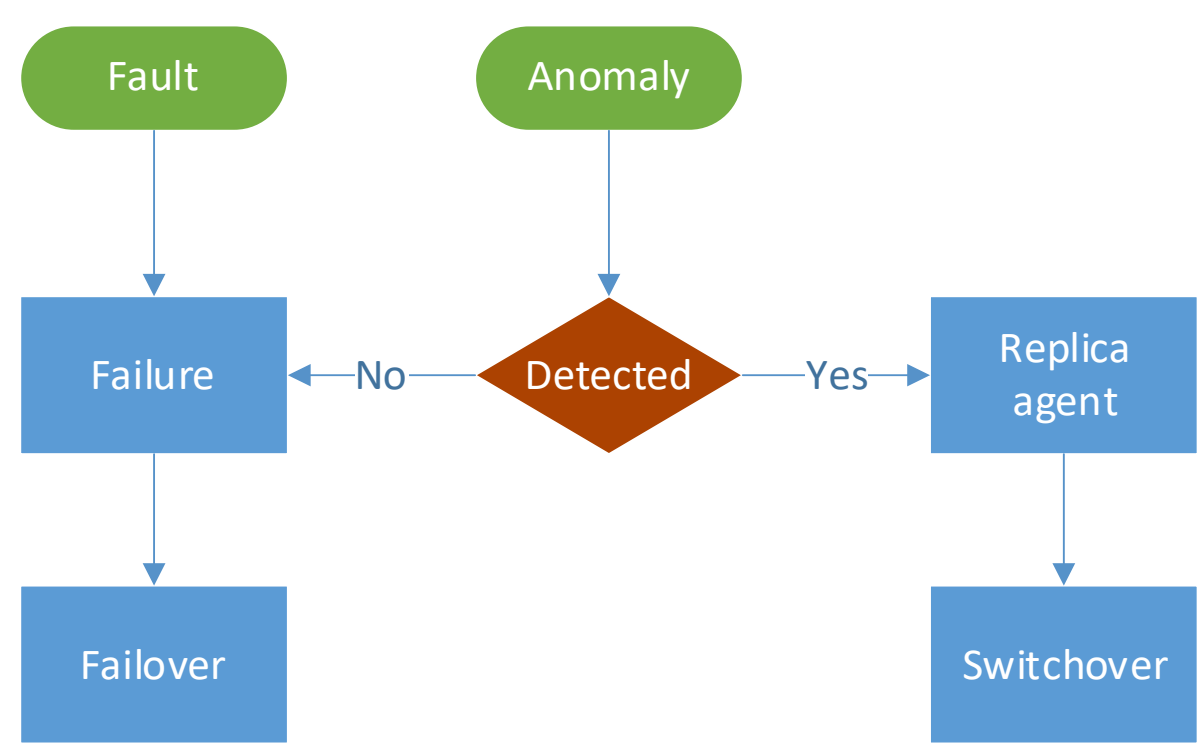

*Fig 3. Service Recovery Paths*

The recovery module supports redundancy that is chosen per microservice, according to criticality and the availability target of that service rather than a single cluster-wide policy.

In line with classical FT design, AMF distinguishes at least two principal models. Under hot standby, a replica runs alongside the active instance and is kept ready to take over with minimal interruption, which favors stringent availability at the cost of continuous resource consumption. Under cold standby, a replica remains inactive or unscheduled until recovery is required, which reduces operational cost but lengthens the time to restore service, including startup and any necessary state preparation. Where no standby is configured, recovery is a restart of the failed instance. Table 1 summarizes the main differences between these models. For stateful services, continuous or on-demand state replication, for example via a state controller, may be required so that switchover or failover does not discard application state, whereas, for stateless services, lighter synchronization may suffice. AMF does not prescribe one model for all CIoT workloads. Instead, it provides the recovery actions, failover after failure and proactive creation or activation of a standby followed by switchover when an anomaly is predicted, so that a software architect can map each service to a redundancy model that matches its allowed downtime, resource budget, and safety role.

*Table 1. Redundancy models in the proposed AMF*

| Aspect | Hot standby | Cold / on-demand standby |
|---|---|---|
| *Standby readiness* | Replica running and ready to take over | Inactive or unscheduled until recovery |
| *Typical recovery trigger* | Failure (failover) and/or predicted anomaly (switchover) | Failure, or anomaly if lead time allows activation |
| *Relative recovery latency* | Lower | Higher (startup / scheduling / sync) |
| *Resource cost* | Higher (continuous consumption) | Lower until activation |
| *State handling* | Continuous or frequent sync when stateful | Sync on demand or at activation when stateful |
| *Typical use* | Mission-critical control paths | Less critical / cost-sensitive services |

In practice, that mapping is a design decision driven by service criticality, i.e., mission-critical control paths typically warrant hot standby, or an equivalent always-ready replica, whereas less critical analytics or auxiliary services may use cold standby. For example, a vehicle-control service responsible for steering, braking, and acceleration may be configured with hot-standby replicas and proactive state synchronization, while an entertainment service with lower safety impact may use cold-standby and fewer replicas. AMF further allows standby activation to be triggered by anomaly detection rather than only by hard failure, so a cold or dormant replica can be brought online during a precursor window when the chosen model and the prediction lead time permit. The experiments conducted in this studyillustrate one such choice, an on-demand standby that is scaled up and switched over under ML-driven recovery for a critical path service, without exhausting the configuration space of the framework. Other deployments may keep an always-on hot pair, place replicas on another node under node-level risk, or combine models across services in the same system.

### *D. State Management*

In AMF, both stateful and stateless microservices are managed through the integration of K8s native controllers with the recovery module's state controller, with the aim of supporting resilient and consistent service availability. For stateless microservices, AMF uses the K8s Deployment controller, which simplifies scaling and recovery by allowing multiple interchangeable replicas to share access to a common Persistent Volume (PV). This arrangement is suitable for services that do not retain internal state across sessions.
Stateful microservices, which maintain service-specific data or context over time, are managed using the K8s StatefulSet controller. Each StatefulSet replica is associated with a dedicated PV, so that state is preserved if the pod is restarted or relocated. This association is important for consistency and data integrity in stateful workloads. However, because the original pod and a replica do not share the same PV by default, state stored on the original PV is not automatically replicated to the replica's PV. As a result, failure of the original pod, or switchover to a replica, may lead to loss of state. To address this issue, AMF adopts the State Controller (SC) proposed by Vayghan et al. [13], which periodically replicates the state of the active pod into the PV of the standby replica so that the two remain aligned. The SC interacts with the K8s API server, monitors cluster events, and responds accordingly. It assigns high-availability roles to active pods and enables them to replicate state data to designated standbys, creating state-replication services as needed. Pods are organized into active–standby pairs, with the standby labeled as a peer and linked to the corresponding active pod. State replication supports both sudden failure and predicted anomaly. Upon sudden failure, the recovery module initiates failover. When the prediction module detects an anomaly in a stateful microservice, the recovery module instead initiates proactive switchover by redirecting traffic to a healthy replica. In either case, the goal is to limit service disruption while allowing the new instance to resume with minimal loss of context. By combining K8s controllers with AMF recovery logic, the framework provides a foundation for availability management across both stateless and stateful components in safety-critical CIoT systems. For application, CPU- or memory-related anomalies, a replica may be placed on the same node. For node-related anomalies, a replica may be placed on another node.

## IV. The Testbed and AMF Deployment

Earlier in [1], [9], we measured E2E service availability on the ToD testbed when pod-related or memory-related faults affect a critical microservice, and showed that K8s' default reactive recovery leaves very high service outage. Those studies characterize the default K8s' recovery, i.e., recovery begins after failure, and service outage duration is governed by container restart and related orchestration delays. As mentioned earlier our goal is to explore that line of investigation by evaluating whether the AMF can improve service availability under a similar class of memory-related faults, by coupling ML-based anomaly detection to proactive recovery rather than relying only on K8s default recovery.

In this section, we introduce the ToD testbed and discuss the AMF deployment on this testbed.

### *A. The ToD Testbed*

The testbed follows our earlier ToD case study [1], a safety-critical CIoT application with an availability requirement of five nines (99.999%) [29]. The system spans three nodes, i.e., an On-Board Unit (OBU), an edge node, and a Remote Driving Station (RDS). Each node runs its own single-node K8s cluster where ToD microservices are deployed on, as shown in Fig 4. For communication between microservices and between nodes, MQTT brokers are deployed on each node, and the brokers form a star topology with the edge broker as the central hub.

The E2E service evaluated in this study is remote speed control. A driver at the RDS sends speed commands that must reach the vehicle's OBU without interruption. RDS speed microservice publishes on MQTT topic set/speed (100 ms period, QoS 0) to the local broker; the message is relayed through the mesh to the edge broker, then to the OBU broker, and finally to OBU speed microservice. Any unexpected outage in that E2E service path directly affects remote control of vehicle speed, so service continuity on this E2E path is the availability metric that matters for the experiment. This E2E service path also has several potential failure points, including the MQTT brokers on each node. In this study, memory-related faults are injected at the edge broker, because it carries all RDS-OBU traffic. The fault model is described in Section IV-D.

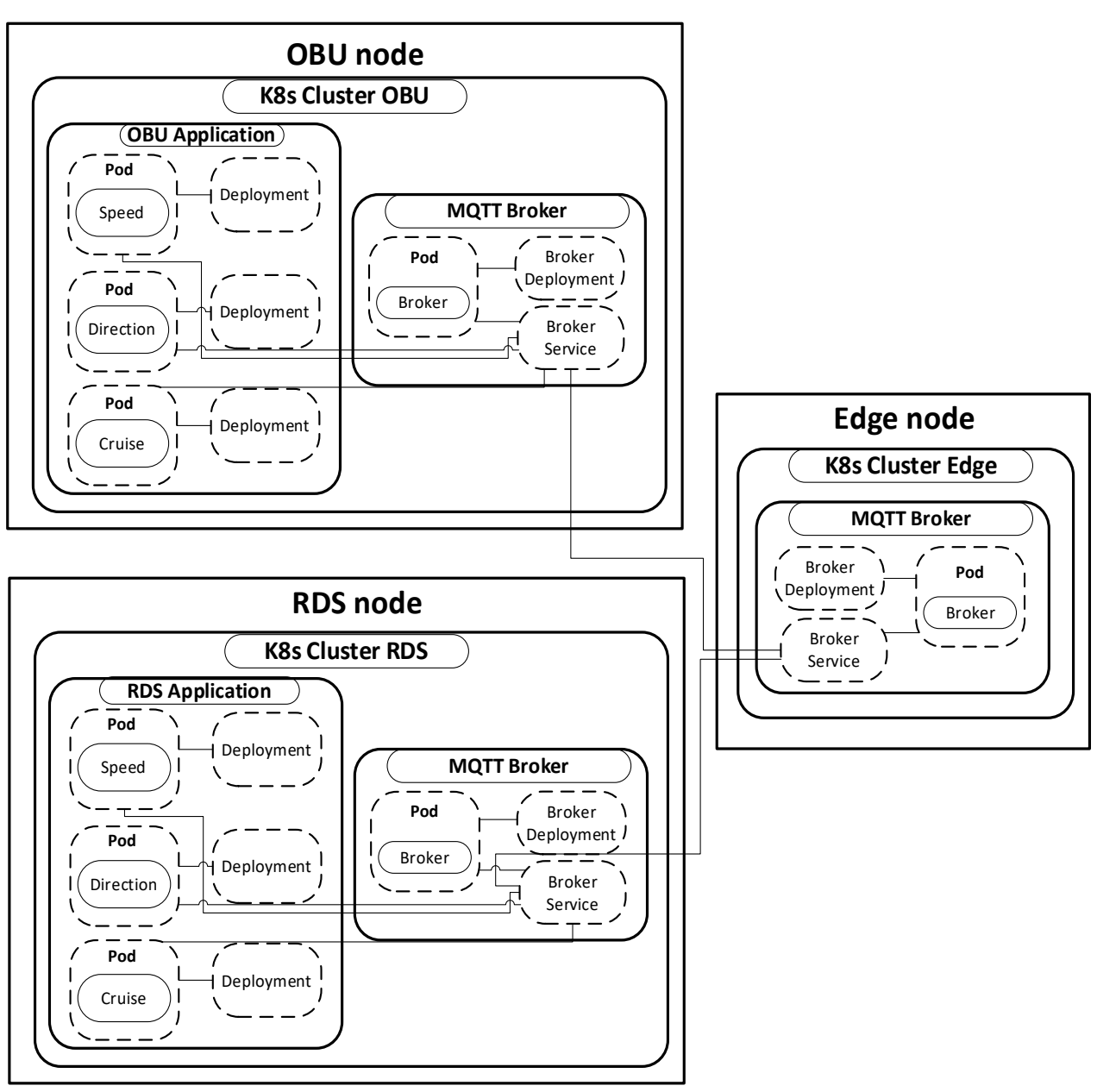

*Fig 4. The ToD Testbed*

Nodes' details of this testbed are summarized in Table 2. The RDS node is an x86 VirtualBox VM, the OBU and edge nodes are Raspberry Pi 4 boards. All three clusters run MicroK8s, and nodes' clocks were NTP-synchronized and logs were correlated in UTC.

*Table 2. Testbed Nodes' Details*

| Node | Type | CPU | RAM | OS | K8s |
|---|---|---|---|---|---|
| RDS | VirtualBox VM x86_64 | 2 × 3.70 GHz | 7.6 GB | Ubuntu 24.04 | MicroK8s 1.34.9 |
| OBU | RPi 4B aarch64 | 4 × 1.80 GHz | 7.6 GB | Ubuntu 24.04 | MicroK8s 1.30.14 |
| Edge | RPi 4B aarch64 | 4 × 1.80 GHz | 7.6 GB | Ubuntu 24.04 | MicroK8s 1.35.6 |

### *B. AMF Deployment on the ToD Testbed*

Beyond the functional design of its modules, AMF must also be mapped onto the distributed topology of a CIoT deployment. This subsection discusses recommended placement of the monitoring, prediction, and recovery modules.

As CIoT systems are typically distributed, the monitoring module is preferably deployed on each node so that microservice health and resource utilization can be observed in real time. Placement of the prediction module is challenging, as this module hosts a trained ML model, there are pros and cons of either having a centralized ML model or training multiple ML models on each node. Therefore, a careful trade-off must be considered. Hosting the model in the cloud may increase the latency of anomaly notifications, which is undesirable in safety-critical settings. In 5G-based CIoT deployments, Multi-Access Edge Computing (MEC) can reduce that latency relative to a remote cloud, yet MEC nodes generally offer fewer resources than cloud platforms, while ML inference and training remain comparatively resource intensive. For our AMF, we recommend deploying the prediction module on the edge node so that predictions can be provided to the other nodes with low latency.

Deployment of the recovery module should follow the application and Cloud-native Network Function (CNF) placement strategy. When microservices run on a container orchestrator such as K8s, the recovery module should match the cluster topology. If K8s is configured as a single multi-node cluster hosting the full CIoT E2E system, the recovery module may be placed centrally on the control plane. If, instead, each site runs as a single-node K8s cluster, the recovery module should be deployed on each node and integrated with the local K8s instance.

For our testbed, as shown in Fig 5, the monitoring module is deployed on all three nodes, prediction module is centralized on the edge node, and recovery module is deployed on all three nodes, consistent with the single-node K8s configuration at each node.

This AMF deployment on the ToD testbed realizes the three AMF modules as a closed data path from telemetry collection through inference to recovery actuation. Monitoring, prediction, and recovery remain functionally separated, i.e., monitoring produces features, prediction produces anomaly scores, and recovery alone mutates cluster state.

For monitoring, resource telemetry is collected with Prometheus [30], exposed on each node via NodePort 30000. A Python client on the corresponding node issues the same six pod-level PromQL queries used to construct the offline training data, that includes six features, i.e., CPU and memory utilization, network and disk I/O throughput, and growth rates derived over five-minute rate and ten-minute deriv windows. More details on this feature set and dataset are discussed in Section V-A.

The prediction module follows a conventional offline-training and online-inference pipeline based on scikit-learn. Offline, a HistGradientBoosting classifier is trained. This ML model is selected based on our robustness analysis presented in Section V. The fitted ML model is exported as a joblib artifact for deployment. Online, an inference process on the edge node polls Prometheus every five seconds, scores each eligible pod, and emits a structured JSON record containing pod identity, anomaly probability, and a binary anomaly flag. No further training or threshold adaptation is performed at runtime. This JSON stream is the sole interface between prediction and recovery.

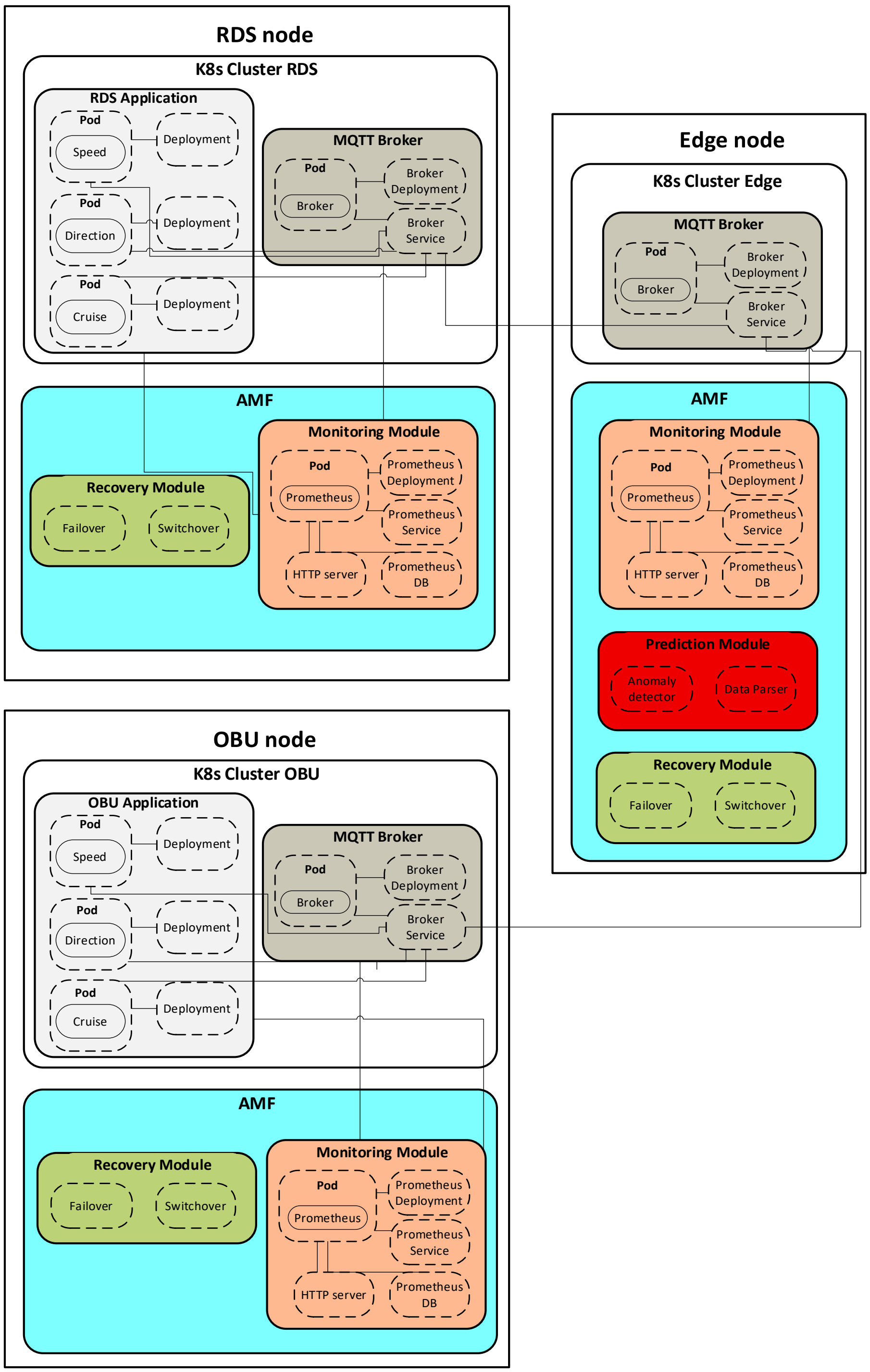

*Fig 5. The AMF deployment on the ToD Testbed*

Recovery module implementation details are provided in Table 3. For failover, K8s deployment controller is used which can restart a pod or container after crash has occurred. For proactive switchover, recovery module uses an on-demand standby broker replica, i.e., the replica creation agent scales up the standby broker deployment, holds overlap, then scales down the primary broker deployment when the prediction module reports an anomaly. Before issuing scale actions, the recovery module applies three actuation gates. The decision threshold is 0.5, i.e, a poll is treated as anomalous only if proba_anomaly ≥ 0.5. Debounce requires three consecutive anomalous polls for the same pod at the 5s interval (about 10s of sustained evidence); any below-threshold sample resets the counter, so an isolated spike cannot trigger recovery. The same three-sample rule is applied again after the standby is Ready, before the primary is retired. The switch hold is a minimum of 15s between the scale-up decision and primary scale-down, so both replicas remain available for a non-zero overlap. The state for the standby bridge configuration is supplied through K8s ConfigMaps with distinct client IDs so the primary and standby can run without clashing.

It is important to note that the prediction module never modifies cluster state, and the recovery module never computes anomaly scores. Consequently, correct detection without this ML pipeline does not shorten the outage. Also, AMF modules do not exchange control messages over the ToD MQTT mesh. In this deployment, MQTT carries the data-plane, i.e., application traffic, whereas AMF needs a control-plane path for anomaly scores and recovery actuation. The evaluated fault targets the edge broker, which is the hub of that mesh, using the same brokers for detector-to-recovery signaling would couple the control-plane to the component under stress and could block actuation precisely when recovery is required. Instead, inference output is passed locally on the edge node, and recovery acts through the local K8s API. Application traffic remains on MQTT but control traffic does not.

*Table 3. The Recovery Module Implementation*

| Component | Trigger | ToD implementation |
|---|---|---|
| Failover | Actual failure (OOM crash) | K8s Deployment controller, reactive container restart |
| Switchover | Detected anomaly (ML) | Proactive standby switchover |
| Replica creation | Anomaly confirmed | Scale K8s deployment 0 → 1 OR 1 → 0 |
| State Controller | Stateful microservices | K8s ConfigMap |
| Actuation gates | Sustained anomaly | Threshold 0.5; debounce 3×5s; switch hold 15s |

## V. ML MODEL SELECTION

The selection of an appropriate ML model is a critical design decision, specifically when balancing detection accuracy against stringent availability requirements. For AMF to close the loop online, the prediction module needs an ML model that is suitable for live deployment. This section summarizes how that ML model was selected. We describe the telemetry dataset, the validation protocols used to compare candidate models, the models under consideration, and a brief robustness analysis that guided the choice of the model used in the experiments of Section VI. A more detailed study of these protocols is reported in [10].

### *A. The Dataset*

The dataset used to train ML model was generated from the ToD testbed through Prometheus metrics collection. It has been made publicly available on GitHub [31] to support future research in safety-critical CIoT. The transition from raw Prometheus telemetry to an analytical dataset requires a structured feature engineering pipeline. We monitor six features, as shown in Table 4, which are the primary metrics at the pod level to characterize system behavior: CPU utilization (%), memory utilization (%), network I/O, disk I/O, CPU growth rate, and memory growth rate. While the first four metrics track instantaneous resource consumption, the growth rates capture the temporal dynamics and workload evolution. These metrics are extracted using the Prometheus query_range API, which ensures that all features are aligned to a normalized time grid. This timestamp synchronization is critical for maintaining consistency across heterogeneous sampling intervals and enabling accurate cross-metric correlation within the ML model.

*Table 4. Feature Set*

| Feature ID | Metric Name | PromQL Queries |
|---|---|---|
| $F_1$ | CPU Utilization (%) | `(sum(rate(container cpu usage seconds total[5m])) by (pod)) * 100` |
| $F_2$ | Memory Utilization (%) | `(sum(container memory usage bytes) by (pod)) / (100 * 1024 * 1024) * 100` |
| $F_3$ | Network I/O (B/sec) | `sum(rate(container network receive bytes total[5m]) + ...[5m]) by (pod)` |
| $F_4$ | Disk I/O (B/sec) | `sum(rate(container fs reads bytes total[5m]) + ...[5m]) by (pod)` |
| $F_5$ | CPU Growth Rate (Units/sec) | `deriv(sum(rate(container cpu usage seconds total[5m])) by (pod) [10m:])` |
| $F_6$ | Memory Growth Rate (B/sec) | `deriv(sum(container memory usage bytes) by (pod) [10m:])` |

To enhance the ML model's ability to detect system instability, the raw metrics are processed using specific temporal windows. A 5-minute sliding window is applied via the PromQL rate function to calculate the per-second average of CPU and I/O metrics, effectively smoothing transient noise that could otherwise trigger false positives. For memory utilization, values are normalized against a 100MB provisioned container limit to represent a consistent percentage of resource exhaustion. Furthermore, we derive higher-order features using the *deriv* function over a 10-minute subquery range. These growth rates serve as first-order derivatives that quantify rapid resource accumulation, which is a primary indicator of anomalies such as memory leaks or CPU thrashing. By combining these smoothed instantaneous values with long-term temporal trends, the feature set provides a comprehensive view of pod health in resource-constrained MicroK8s environments. Prometheus Metrics and Corresponding PromQL Queries are presented in Table 4 and the specific parameters used during the data acquisition phase are summarized in Table 5. These settings were chosen to ensure high-granularity monitoring while maintaining the stability of the resource-constrained MicroK8s environment. The resource boundaries for each pod were strictly defined with a memory limit of 100 MB and a CPU limit of 500m (0.5 cores). This constrained environment is intentional, as it allows the anomaly detection model to be trained on realistic resource-pressure scenarios common in IoT and edge environments. To capture rapid fluctuations in system behavior, the Prometheus scraping interval was set to 1 second, providing a high-resolution raw data stream.

For the final dataset construction, data was captured across two non-contiguous120-minutes windows (totaling four hours of observational data) to ensure a comprehensive representation of both nominal and anomalous system states. The first interval, designated as $D_{nominal}$, captures the system under standard operational workloads. The second interval, $D_{stressed}$, encompasses a series of high-intensity resource stressors and synthetic faults. While the raw data was scraped at a high-granularity 1-second interval, the datasets were down sampled using a 5-second step window. This 5-fold temporal aggregation serves as a primary denoising mechanism, mitigating high-frequency transient spikes and reducing computational overhead without compromising the resolution required to identify the 5-minute and 10-minute trends critical for our feature engineering pipeline. This specific step interval ensures that the resulting features maintain the necessary sensitivity to capture the gradual degradation preceding system failure.

*Table 5. Data Collection Settings*

| Setting | Value |
|---|---|
| Pod MEM Limit | 100 MB |
| Pod CPU Limit | 500m |
| Prometheus Scraping Interval | 1 second |
| Data Collection Duration | 2 × 120 min (non-contiguous; 4 h total) |
| Step Interval | 5 seconds |

For $D_{stressed}$, a controlled anomaly injection mechanism was implemented using a custom Bash script and the stress-ng utility [32]. This script executes a sequence of synthetic workloads directly within the target pods via microk8s kubectl exec, allowing for the precise simulation of resource contention in a containerized environment. To ensure a clear distinction between nominal and faulty states, the process begins with a 120-second Baseline phase to record idle system behavior. This is followed by a Steady High Load phase, maintaining CPU at approximately 80% and memory at 60 MB for 120 seconds, and two subsequent Ramp phases (CPU and Memory) that incrementally increase resource pressure in 60-second intervals to capture the transition from healthy to degraded states. Summary of anomaly injection profile and phases for this dataset collection is presented in Table 6. The injection sequence also incorporates highly dynamic patterns to test the model's sensitivity to transient faults. This includes a Burst Spikes phase, consisting of three 20-seconds 90% CPU bursts, and an Oscillation phase that toggles the CPU between 80% and 20% every 30 seconds. The entire sequence spans approximately 18 minutes of active fault injection per cycle. To facilitate accurate labeling during the supervised learning process, the script generates timestamped logs for the start and end of each phase. These logs allow for the seamless alignment of the ground-truth anomaly events with the 5-second Prometheus telemetry stream, ensuring the dataset accurately represents the physical reality of the system under stress. Fig 6 shows a plot of dataset for microservice speed pod on OBU node. This 4-hour dataset contains 2-hours normal operation under no extra load ($D_{nominal}$ from 01:45:00 to 03:45:00), and 2-hours stress period ($D_{stressed}$ from 07:02:00 to 09:02:00). It can be observed that CPU percent and memory percent have clear spikes during stress period, and CPU growth rate and memory growth rate are fluctuating. It can also be observed that, in $D_{nominal}$, memory percent is constantly ~100%. This is the single-feature anomaly in only one pod and this latent anomaly provides a rigorous baseline. It is important to note that the plot in Fig 6 only shows data for one pod on one node, whereas the whole dataset contains data from 9 pods in total on 3 nodes. Plots for whole dataset are not presented in this paper for brevity.

*Table 6. Anomaly Injection Profile for Dataset Collection*

| Phase | Duration | Resource Pattern |
|---|---|---|
| Baseline | 120 sec | Idle / Nominal behavior |
| Steady High | 120 sec | ~80% CPU + 60MB Memory |
| CPU Ramp | 240 sec | 20% → 40% → 60% → 80% |
| Memory Ramp | 240 sec | 20MB → 40MB → 60MB → 80MB |
| Burst Spikes | 180 sec | 3 cycles of 90% CPU spikes (20 sec load, 40 sec rest) |
| Oscillation | 180 sec | 3 cycles of 80% / 20% CPU toggle (30 sec each) |

For ML model selection, we considered both supervised and unsupervised learning approaches. While unsupervised methods such as K-Means or Isolation Forest offer the advantage of operating without labeled data, they present significant challenges in automated recovery loops. In such models, the resulting clusters often require manual post-facto interpretation to distinguish between specific fault types, hindering real-time response. Furthermore, in safety-critical domains, explainability is paramount. Therefore, we selected three supervised ML models, Random Forest (RF) [33], Logistic Regression (LR) [34], and Hist-Gradient Boosting Classifier (HGBC) [35]. These models not only facilitate feature-level validation to identify the specific drivers of an anomaly score but are also naturally suited to capture the cliff-edge effects (e.g., sudden CPU thrashing or memory exhaustion) common in microservice resource contention.

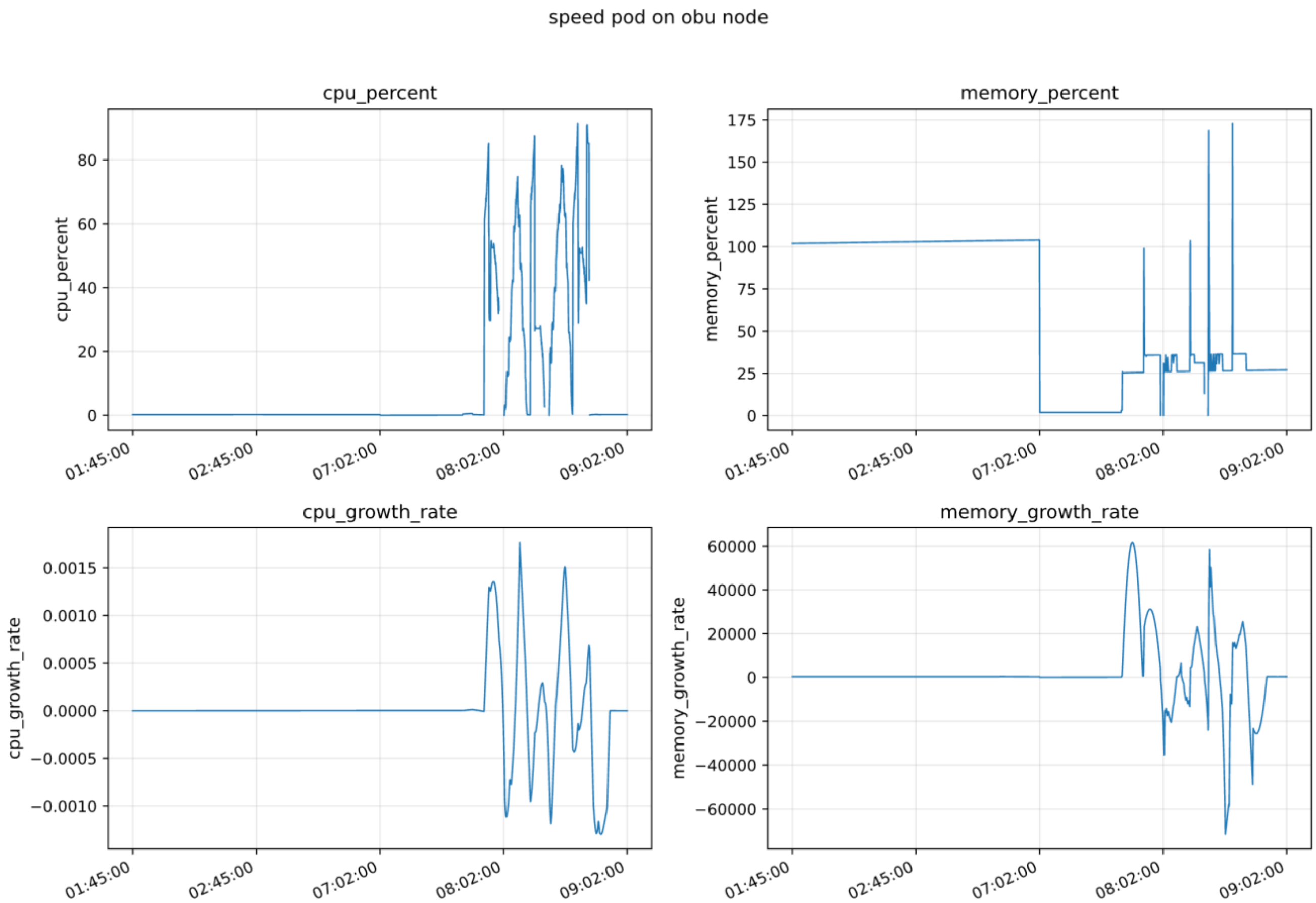

*Fig 6. Dataset plot - One pod on one node [10]*

### *B. Validation Protocols*

Candidate detectors are compared under three complementary validation protocols, inspired by [36], that reflect different relationships between the data used for training and the data used for evaluation. The protocols are used here to select a model suitable for deployment, not as a standalone study of open-world Out-Of-Distribution (OOD) learning. Taken together, they contrast an optimistic in-distribution baseline with more realistic operating conditions for safety-critical CIoT systems. In this article, these protocols guide ML model selection for the live AMF loop. The model deployed in Section VI is chosen on the basis of P1, which balances offline evidence of recovery under stressed conditions with the availability of limited labeled support.

a) *P0 – i.i.d. Baseline:* Models are trained and tested on a shuffled, stratified split of the consolidated dataset (containing both $D_{nominal}$ and $D_{stressed}$). Training and test samples are drawn from the same distribution, approximating an independent and identically distributed (*i.i.d.)* scenario. P0 therefore provides an optimistic upper bound on classifier performance when the evaluation distribution matches the training mixture. Telemetry is temporally autocorrelated, so P0 should not be interpreted as a strict *i.i.d.* guarantee.

b) *P1 – Few-Shot Domain Adaptation:* P1 represents a practical deployment setting in which some labeled stressed-regime data are available for model training. Training uses $D_{nominal}$ together with a stratified 30% support subset of $D_{stressed}$; evaluation uses the held-out 70% of $D_{stressed}$. This protocol measures how well detection performance recovers when the detector is retrained on the concatenated labeled support, rather than classical few-shot adaptation in the machine-learning sense.

c) *P2 – OOD Generalization:* P2 trains on the full $D_{nominal}$ data (both classes) and evaluates on the full $D_{stressed}$ data, with no labeled target-regime support at training time. It therefore assesses transfer across operating regimes on the same testbed. P2 is a zero-shot regime-transfer setting, not an OOD benchmark across unrelated deployments.

### *C. ML Model Robustness Analysis*

Offline robustness analysis is used here to choose an ML model for the AMF recovery loop, not to claim a new learning method. In AMF, an anomaly score is not an end in itself, in fact, a positive decision can trigger standby creation and switchover. The relevant trade-off is therefore between false alarms and missed precursors. Precision measures how often a raised anomaly corresponds to a true anomalous sample, i.e., low precision implies frequent false alarms and unnecessary recovery actions. On the other hand, recall measures how often true anomalies are detected, i.e., low recall implies missed precursors and lost opportunity to act before failure. For closed-loop availability management, both matter, but they are not interchangeable, i.e., a high-recall, low-precision detector is a poor driver for automatic switchover, whereas a high-precision, low-recall detector may rarely act when action would help. Other evaluation metrics, accuracy, F1-score and AUC are reported in [10] and are omitted here, because they do not directly affect the recovery action trade-off.

Candidate models, RF, LR, and HGBC, are compared under protocols P0-P2. Table 7 reports precision and recall under P0, P1 and P2, mean ± standard deviation over five random seeds (42-46). Full multi-metric results can be found in [10].

*Table 7. Precision and Recall under Protocols (mean ± std, 5 seeds)*

| ML Model | Precision | | | Recall | | |
|---|---|---|---|---|---|---|
| | ***P0*** | ***P1*** | ***P2*** | ***P0*** | ***P1*** | ***P2*** |
| RF | 1.000± 0.000 | 0.999± 0.001 | 1.000± 0.000 | 1.000± 0.000 | 1.000± 0.000 | 0.353± 0.005 |
| LR | 0.897± 0.005 | 0.845± 0.006 | 0.403± 0.000 | 0.916± 0.002 | 0.877± 0.005 | 0.947± 0.000 |
| HGBC | 0.995± 0.002 | 0.991± 0.007 | 1.000± 0.000 | 0.996± 0.002 | 0.989± 0.005 | 0.465± 0.000 |

Under P0, both tree ensembles achieve near-perfect precision and recall. This result is an optimistic upper bound when train and test samples are drawn from the same pooled mixture; it should not be read as evidence that the same scores will hold after deployment.

Under P2, the precision–recall trade-off becomes decisive for AMF. RF and HGBC keep precision at 1.000, so almost every positive prediction is a true anomaly, but recall falls to 0.353±0.005 and 0.465±0.000, respectively, i.e., most stressed-regime anomalies are missed. LR shows the opposite pattern, with high recall (0.947) but precision of only 0.403. In an AMF setting where positives can drive switchover, that low precision implies a large false-alarm rate and is therefore unsuitable for live recovery actuation, despite strong recall.

Under P1, limited labeled stressed-regime support restores strong precision and recall for the tree ensembles, i.e., RF and HGBC both remain over 0.99. LR also improves relative to P2 but remains weaker than the ensembles on precision. Protocol P1 is therefore the training regime we adopt for deployment, as it retains high precision while recovering the recall needed to detect precursors under stressed conditions.

ML model selection for the AMF prioritizes (i) high precision, so recovery is not dominated by false alarms, and (ii) among high-precision models, higher recall under a change of operating regime, so precursors are missed less often. LR is discarded for live actuation because of its low P2 precision. RF and HGBC both satisfy the precision requirement under P1 and P2. Between them, HGBC is selected because it retains higher recall under P2 (0.465 vs 0.353) while preserving precision of 1.000. Under P1, RF is marginally stronger on precision and recall, we do not claim HGBC is uniquely optimal on P1 alone. The live detector used in our deployed AMF is therefore the P1-trained HGBC model with decision threshold of 0.5. A more complete robustness study, including accuracy, F1-score, and AUC, is reported in [10].

## VI. Availability Experiments and Results

This section reports on the E2E SOT for the remote speed-control path under a staged memory anomaly and a fault profile discussed hereafter. We discuss results under default K8s recovery, i.e., without the proposed AMF (or AMF disabled) . We then present results with the AMF deployed, and the anomaly detector is coupled to on-demand standby switchover. The experiments use the same testbed, the same injection target, i.e., edge broker, and the same OBU speed service path measurement.

### *A. Evaluation Metrics*

The primary availability metric in this study is E2E Service Outage Time (E2E SOT) on the speed control service path. An E2E service is the application function delivered from a source microservice to a consumer microservice. In the ToD testbed, RDS speed microservice originates speed commands for the remote driver, and OBU speed microservice receives those commands for the vehicle side. E2E SOT, as illustrated in Fig 7, is the time interval ($t_3 - t_1$) during which the service is not delivered to the consumer after a failure, i.e., the source may still be publishing, and individual pods or brokers may later report healthy, but the consumer does not receive the command stream in continuity. Fig 7 also shows the proactive path: if an anomaly at $t_0$ is detected and switchover $R_s$ completes before failure, $t_1$ may not occur, Section VI-D evaluates that case.

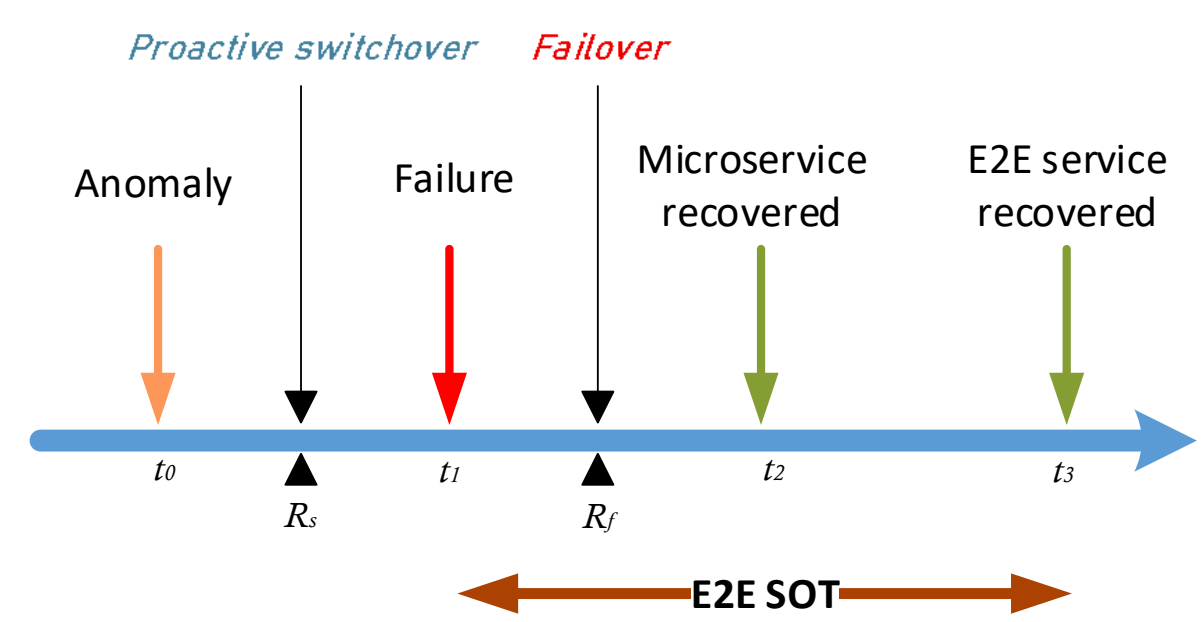

*Fig 7. Events Relationship and Evaluation Metric*

E2E SOT is measured on the OBU using a continuous subscriber to MQTT. The RDS publisher embeds its clock time in each payload and publishes every 100 ms at QoS 0. The OBU client compares consecutive payload timestamps in the received sequence, so a large gap means messages missing from that sequence, not merely delayed delivery of a message that still arrives, and the gap does not depend on RDS-OBU clock offset. This definition follows our earlier ToD availability work [1], [9], adapted here for the AMF experiments. Furthermore, a sample is recorded only when the difference exceeds 250 ms, where an interval with no recorded sample means no gap above 250 ms, i.e., an upper bound of 0.25 s, not a proven zero. Because the healthy QoS 0 path is not loss-free, we characterize background variation from no-stress windows. For a failure or recovery event, we report the largest recorded gap within 30s of that event, so later ordinary variation is not attributed to it. Under default K8s recovery, that quantity is E2E SOT after a realized fault. Under AMF switchover, we report the same gap measure at cut-over and compare it to the healthy baseline.

It is also worth noting the difference between a recovered microservice and recovered E2E service, i.e., $t_2$ and $t_3$

respectively from Fig 7. A recovered microservice doesn't necessarily mean that E2E service is also recovered, as after microservice recovery, E2E service path may take more time to update system state, e.g., load system configuration. This differentiates service availability from system availability. For instance, in the ToD testbed, a broker pod can return to Ready state and K8s can report recovery complete, while brokers' state, i.e., broker bridge configuration, has not yet restored the communication flow. E2E SOT therefore captures outages visible to the application consumer. Results in Section VI report E2E SOT under default K8s recovery and under AMF-driven switchover on the same E2E service path.

### *B. Faults & Anomalies Profile*

In safety-critical systems, harmful conditions often develop gradually, i.e., performance degrades, internal state becomes unstable, and only later does a failure interrupt the service offered to the user. Availability management is most useful in that interval, when a proactive recovery action can still prevent or shorten an outage. This subsection defines how anomalies and faults are modeled on the testbed and the anomaly/fault injection target.

An anomaly is an abnormal behavior that signals risk but does not by itself stop the E2E service, e.g., sustained high resource usage while the component remains operational. Whereas a failure is an event that interrupts service until recovery is completed. An anomaly may precede a failure, detecting an anomaly does not mean a failure has already occurred. E2E SOT is tied to the failure's impact on the consumer, not to the first anomalous event, as illustrated in Fig 7. An anomaly occurs at $t_0$, and if it is detected and proactive switchover ($R_s$) is applied, then a failure can be avoided. But if either an anomaly is missed or $R_s$ is not applied, then it can become a fault and cause a failure ($t_1$) which would need a failover ($R_f$) to recover the service.

The anomaly and fault profile here follows memory-related stress characterized in [9]. It uses a staged injector on the edge broker pod, as described in Table 8. Phase 1 aligns with sub-limit stress used in offline training data, as discussed in Section V, and allows Prometheus rate features to update. Phase 1 also allows to record gaps/jitter on the service path during a short no-stress window, so that cut-over or failure gaps can be compared to the immediately preceding healthy baseline rather than to an abstract bound alone. Phase 2 matches broker OOM behavior in [9]. Phases run consecutively with no idle gap between them. An event with only sub-limit stress can be detectable yet may not produce a clear outage, whereas an event with immediate OOM leaves almost no precursor window. The staged profile supports both precursor detection and, when recovery does not act in time, an E2E service outage. The injector does not execute recovery actions.

*Table 8. Fault and Anomaly Profile*

| Phase | Load | Duration | Role |
|---|---|---|---|
| 0 | None | 60s | Baseline |
| 1 | MEM and CPU stress | 120s | Anomaly |
| 2 | Over-limit MEM | 60s | Failure |

Stress was applied at one microservice, the edge MQTT broker, see Fig 5, because it is the central hub of the broker mesh for the evaluated service path, i.e., every RDS-to-OBU message for remote speed control passes through it. Failure or loss of that relay therefore affects the full E2E service, not a single local consumer or publisher in isolation. Injecting the same profile on OBU or RDS speed microservice would test failure on one leg of the path, a different failure semantics question. Injecting a non-hub broker would not reproduce hub-wide loss of connectivity for this service. The campaign thus exercises availability management on the highest-impact single point along the defined E2E service path, which is sufficient to validate whether precursor detection coupled to recovery improves continuity on that path. AMF is intended to be configured per microservice, therefore demonstrating the closed loop on this central broker hub role establishes the mechanism without repeating the same study on every other pod, which would mainly vary component type rather than the hub-outage hypothesis under test.

### *C. E2E Service Outage without AMF*

We first measure E2E SOT when AMF-driven switchover is disabled and recovery is left to default K8s recovery, that is, in-place container restart after OOM. Injection follows the staged anomaly-to-fault profile on the edge MQTT broker. E2E SOT is measured at the OBU subscriber on set/speed as defined earlier in this section.

Five consecutive injections were executed on the same primary pod. Phase 1 completed without OOM in every run, whereas Phase 2 always caused an OOM crash. Because the runs were spaced more closely than the K8s backoff-reset interval, restart delay accumulated across the series. Fig 8 shows E2E SOT for these five runs. On the first crash, E2E SOT is 2.67s. That first-crash outage is the fair baseline for comparison with AMF, i.e., it is consistent with earlier lab measurements of reactive recovery on this ToD testbed [9]. Runs 2–5 show that repeated failure of the same container inflates E2E SOT under K8s exponential restart backoff, reaching 83.31s by the fifth crash (see Fig 8). This exponential growth is a property of reactive failover under repeated failures.

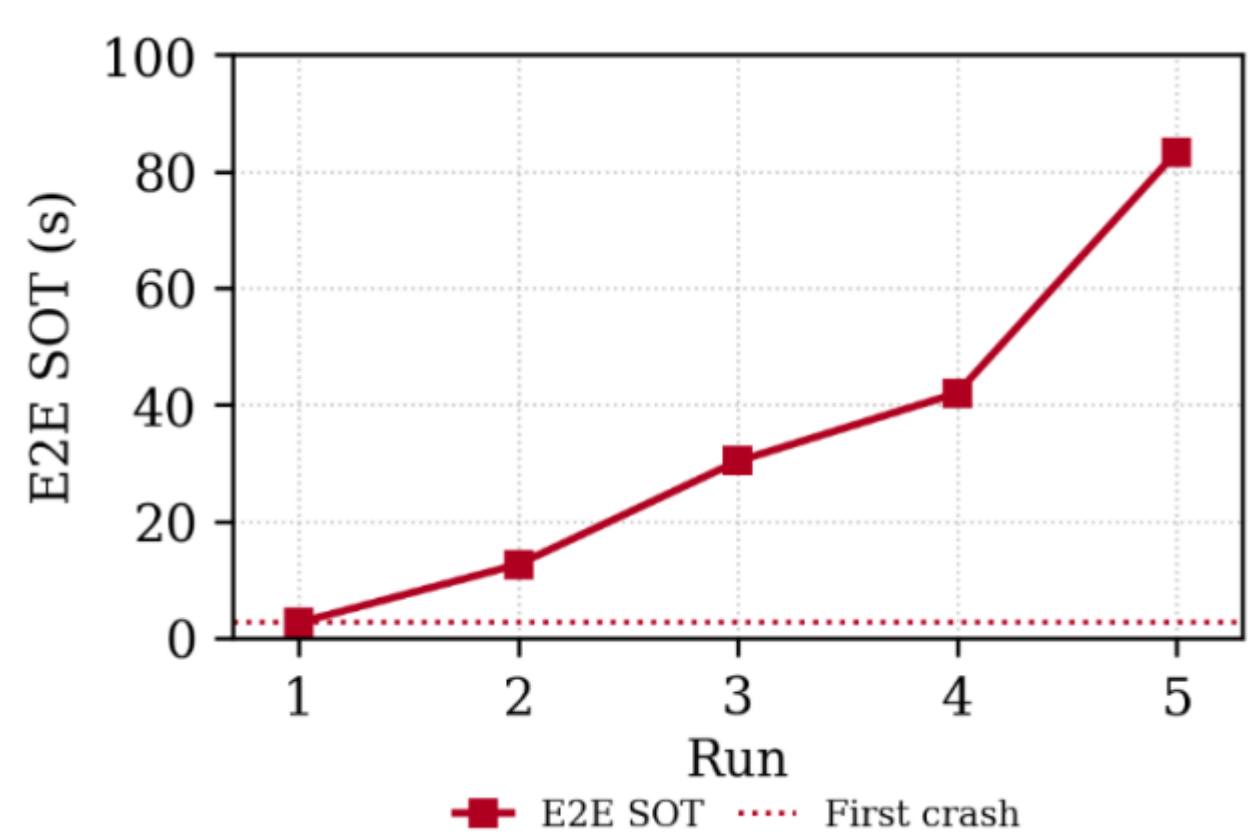

*Fig 8. E2E SOT without AMF*

To separate detection from recovery, we also ran one injection with the anomaly detector in observe-only mode and the recovery agent disabled. Detection latency is the time from Phase 1 start to the first anomaly flag; lead time is the time from that flag to OOM. In this run, detection latency was 14s (proba_anomaly ≈ 0.999) and lead time about 107s, with no false positives in the immediately preceding healthy window. The 14s latency is expected for this instance, as the detector polls every 5s, and the PromQL rate and growth features need a short interval after stress begins before the score crosses the threshold. It is therefore not a model-compute delay of 14s, but the reaction time of the monitoring and inference loop. E2E SOT was still 4.41s, a multi-second, crash-aligned hole of the same class as the 2.67s first-crash baseline in Fig 8. The difference between these two first-crash runs is consistent with ordinary variation in reactive restart and mesh reconnection, i.e., the detector did not actuate recovery.

*D. E2E Service Outage with AMF*

We now enable the full AMF loop, i.e., the anomaly detector output is consumed by the recovery module, which performs on-demand standby scale-up followed by switchover. Five injections were executed where injection target, fault profile, and E2E SOT measurement are unchanged from previous subsection VI-C. In all five runs the replica creation agent acted while the primary broker pod was still Running and serving. It confirmed the anomaly 18-27 s after the onset of Phase 1, at proba_anomaly ≈ 0.999 in every run, and issued the standby scale-up. This 18-27 s interval is not model compute time, it is the observe-only detection latency of Section VI-C (approximately 14s for the monitoring and inference loop to cross threshold) plus the confirmation debounce, which requires three consecutive above-threshold samples at the 5s polling interval before any action is taken. The agent then waited until the standby Deployment reported Ready and only afterwards scaled the primary to zero, so both replicas were Ready simultaneously for a non-zero interval and there was no window in which neither broker was serving. The interval from the scale-up decision to retirement of the primary was 15-16 s across the five runs, set by the switch-hold parameter of Section IV, this bounds the dual-Ready overlap from above, because standby rollout time within that interval was not separately instrumented. Switchover completed 34-42 s after the onset of Phase 1, i.e., 78-86 s before the scheduled escalation at the end of the 120s anomaly phase.

Because the primary broker pod was retired mid-Phase 1, the injected fault never materialized on that pod. The anomaly load ended with the pod shortly after each switchover, and the escalation command could not be delivered against a target that no longer existed. No OOM occurred and no container restart was recorded.

Table 9 reports, for each run, the largest outage, we call it jitter, on the service path in the healthy (no-stress) window immediately preceding injection, and the largest E2E SOT, in the 30s after the first switchover. In runs 2 and 4, the OBU client recorded no above-threshold inter-message gap near switchover. Under the 250ms recording rule, those two runs are upper bounds rather than measured zeros, and we report them as < 0.25s. The probe was live across both windows, samples tens of seconds before and after those switchovers.

The E2E SOT falls inside the jitter of healthy mesh. Across no-stress windows in the campaign, jitter have a median of 0.77s and a maximum of 1.09s, whereas in the windows following the five switchovers the median is 0.68s and the maximum is 1.05s. The mean E2E SOT with AMF is 0.61s which is smaller than the mean jitter i.e., 0.99s. We therefore cannot claim zero downtime. We claim that AMF-driven switchover produced no interruption larger than the jitter of the healthy path on this testbed, in contrast to the 2.67s E2E SOT without AMF in Fig 8.

*Table 9. E2E SOT with the AMF*

| Run | Jitter (s) (no-stress baseline) | E2E SOT (s) |
|---|---|---|
| 1 | 1.01 | 0.99 |
| 2 | 1.01 | < 0.25 |
| 3 | 0.95 | 0.50 |
| 4 | 1.00 | < 0.25 |
| 5 | 0.98 | 1.05 |

Fig 9shows E2E SOT with the AMF. Open markers denote runs with no gap above the 250 ms recording threshold, plotted as 0.25 s. The dashed curve is the healthy-path jitter for those runs, the maximum outage in the no-stress window immediately before each injection, and the dotted line is the first-crash E2E SOT without AMF, from Fig 8. With the AMF, E2E SOT stays below that first crash on every run. On four of the five runs it also stays below the jitter curve, and on the fifth it is 1.05 s, slightly above the jitter for that run. From Fig 8 and Fig 9, it can also be observed from that how AMF avoids the exponential rise in E2E SOT without AMF. The injection, the fault profile, and the E2E SOT measurement are the same in both experiments. Without AMF, the same pod is restarted in place after each OOM, and the outage grows exponentially from one crash to the next. With AMF, switchover removes the primary during the anomaly, before the OOM, and each run in Fig 9 begins from a fresh primary. The E2E service therefore does not remain on a pod whose recovery time is still increasing.

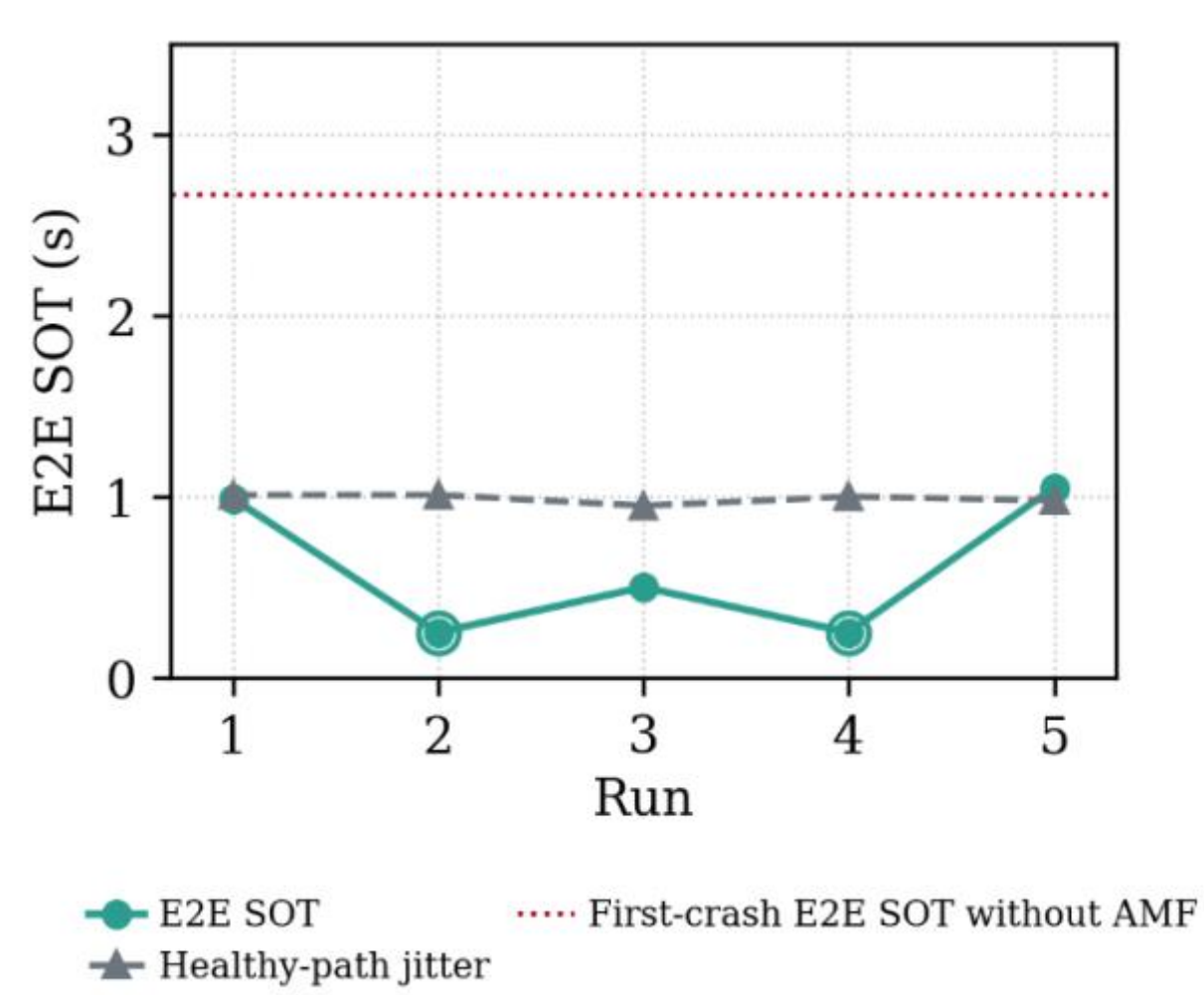

*Fig 9. E2E SOT with the AMF*

## VII. Discussion and Threats to Validity

The proposed AMF is a generic availability management loop, i.e., monitor, predict, recover, that can sit beside reactive orchestration on microservice platforms. The AMF deployment on the ToD testbed, as discussed in Sections IV–VI, is one closed-loop instance and not the architecture's definition.

The architectural contrast with default K8s recovery is therefore between when remediation is allowed to start. Failover ($R_f$ in Fig 7) begins after failure, whereas switchover ($R_s$) begins during the anomaly window while the primary is still serving. On the measured path, first-crash E2E SOT under failover is 2.67 s (detect-only 4.41 s), whereas AMF switchover completes in Phase 1 with a maximum recorded payload gap of 1.05 s at cut-over and no OOM on the retired primary. Treating the two recordings below the 250 ms threshold as 0.25 s for summarization gives a mean maximum gap of 0.61±0.39 s under AMF, inside the healthy-path jitter of this mesh (mean 0.99±0.03 s). Those cut-over measurements are therefore not classified as failover-class E2E SOT. Break-before-make laboratory restorations on the same probe produce E2E SOT of 2.36-3.93 s (mean 3.12 s) and confirm that multi-second outages remain visible when ordering is reversed. The exponential growth in Section VI-C, from 2.67 s on the first crash to 83.31 s on the fifth, is repeated in-place restart of one pod. Those crashes came too close together for the restart wait to clear, so E2E SOT grew from one crash to the next. The later crashes show that growth. The proposed AMF is compared with the first crash, under the same injection, fault profile, and E2E SOT measurement. AMF also helps avoid the exponential growth on its own runs because switchover removes the primary during the anomaly, before the OOM, and each AMF run begins from a fresh primary.

AMF can improve E2E availability only when the anomaly lasts long enough for detection and gated recovery to finish before the fault. That requirement follows from the monitor-predict-recover loop in Section III, not from this testbed alone. Any instantiation of the AMF needs an anomaly window longer than detection latency plus the recovery module's gates (threshold, debounce, switch hold). Abrupt faults with almost no anomaly window fall back to reactive failover by construction, so do missed detections if the live regime shifts toward the low-recall P2 case of Section V. The staged profile in Section VI exercises the case the architecture targets, i.e., a usable anomaly window before fault. How long that window is in a given deployment depends on the fault and the metrics pipeline, it is not a promise that every memory related failure warns for two minutes.

The five-nines (99.999%) availability requirement cited for the ToD testbed [29] must be read against the same separation of concerns. This campaign does not measure long-term availability against that target, it rather measures second-scale E2E SOT on one path under a controlled fault. In the present instance, AMF finished switchover with tens of seconds of Phase 1 remaining, so restart- and backoff-dominated outage is removed from the critical path. Residual recorded gaps at cut-over are then on the order of the healthy-path jitter of a non-real-time, QoS 0 broker mesh on constrained nodes, not the product of a late recovery decision. That jitter is a limitation of the present testbed and service path, not of the AMF loop. Whether a given deployment ultimately meets five-nines depends on path timing, fault rate, and observation horizon beyond a single staged injection, which these experiments do not claim to establish.

Several instance limits follow without narrowing the architecture itself. The evaluated recovery path is the edge-co-located subset of Section III, with a single switchover per configuration and no role exchange afterward. No spurious switchover appeared in the immediate pre-injection healthy windows of the five AMF runs; quantifying false-alarm cost under long healthy operation remains open. Extending the loop across tiers, fault classes, and control-plane scales is potential future work.

These limits set the scope of the claim, and the comparison inside them is still worth reporting. The observations concern the edge broker, through which every message on this service path passes. The AMF and the default K8s recovery were observed under the same injection, the same staged memory and CPU stress, and the same live OBU measurement of E2E SOT. Although five AMF runs do not estimate an E2E SOT distribution for other CIoT systems, we believe they are enough to support the conclusion for this testbed.

What another ToD service or another CIoT system may take from this work is the condition, not the measured seconds. Where a usable anomaly window exists and standby switchover can finish before the fault, AMF has the potential to keep that multi-second outage off the E2E service path. The values reported here belong to this broker, this probe, and this stress profile. A fault that reaches OOM before detection, gating, and switchover completion follows default K8s recovery, as the architecture already specifies.

## VIII. Conclusion

This article proposed an AMF, a generic availability management framework that couples resource monitoring and ML-based anomaly detection with proactive recovery for microservice-based safety-critical CIoT systems. Building on a prior architectural proposal [11], we described the monitor-predict-recover loop, selected an anomaly detector using offline protocols that include a change of operating regime [10], and evaluated one closed-loop instance on a three-tier tele-operated driving testbed under a staged memory and CPU exhaustion scenario.

The experimental results clarify what the architecture contributes. Default K8s recovery leaves a first-crash E2E SOT of 2.67s and, under repeated in-place restarts, E2E SOT of tens of seconds once restart backoff accumulates. With the proposed AMF, switchover completes during the anomaly window, i.e., across five runs, the maximum recorded E2E SOT at switchover is at most 1.05s (within healthy-path jitter), the injected fault never materializes as an OOM on the retired primary pod, so the backoff-dominated regime is never entered. Furthermore, the healthy-path jitter is a property of the present testbed and service path, not evidence that the AMF decision was late, i.e., switchover finished with tens of seconds of the anomaly window still unused. Meeting the five-nines availability requirement remains a joint outcome of timely availability management, service path latency, fault rate, and observation horizon beyond a single staged injection. This evaluation shows that the proposed AMF can complete switchover in time to avoid failover-class E2E SOT, it does not show that the present testbed and service path is free of healthy-path jitter. The proposed AMF can complement reactive orchestration when coupled with proactive detection and recovery.

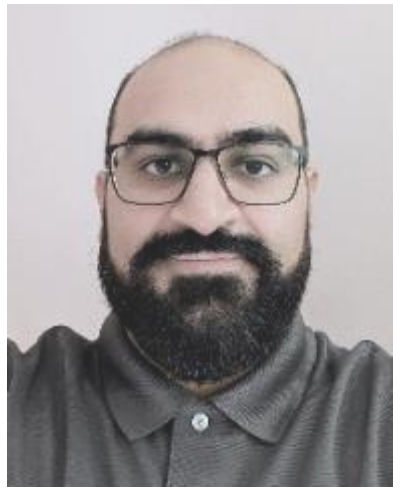

**Hassaan Siddiqui** is currently pursuing the Ph.D. degree in Electrical and Computer Engineering at Concordia University, Montreal, Canada. He brings over 15 years of industrial experience in cloud-native infrastructure, platform engineering, and DevOps, currently serving as a Senior DevOps Engineer at Tecsys Inc., Montreal, Canada. His prior industry background includes engineering roles at Ericsson. His research interests lie at the intersection of Applied AI and software reliability, focusing on microservices, anomaly detection, Kubernetes observability, Cloud, and IoT systems.

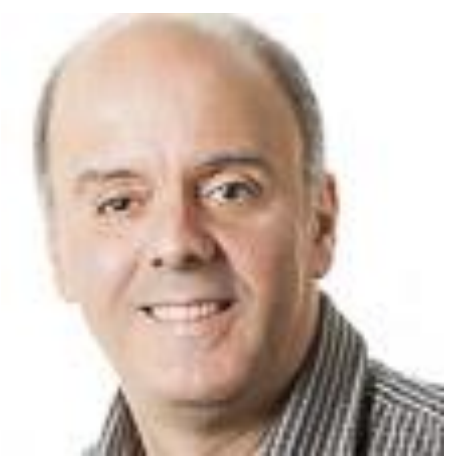

**Ferhat Khendek** (Member, IEEE), received the Ph.D. degree from the Université de Montreal, Canada. He is a Professor with the Department of Electrical and Computer Engineering, Concordia University, where he held the NSERC/Ericsson Senior Industrial Research Chair in Model Based Management from 2011 to 2022. His research interests are in model-based software engineering and management, cloud computing, network function virtualization, IoT, dependability, and software systems validation and testing.